\documentclass[twocolumn]{aastex631}
\usepackage{dcolumn}

\newcommand{\gr}{$\gamma$-ray}
\newcommand{\fermi}{{\it Fermi}}

\shorttitle{$\gamma$-ray pulsations of PSR J1835$-$3259B in NGC 6652}
\shortauthors{Zhang et al.}
\graphicspath{{./}{figures/}}

\begin{document}

\title{Discovery of $\gamma$-ray pulsations of PSR J1835$-$3259B in the Globular Cluster NGC 6652}

\author{Pengfei Zhang}
\affiliation{Department of Astronomy, School of Physics and Astronomy, Key Laboratory of Astroparticle Physics of Yunnan Province, Yunnan University, Kunming 650091, People's Republic of China; zhangpengfei@ynu.edu.cn; wangzx20@ynu.edu.cn}

\author{Yi Xing}
\affiliation{Key Laboratory for Research in Galaxies and Cosmology, Shanghai Astronomical Observatory, Chinese Academy of Sciences, 80 Nandan Road, Shanghai 200030, People's Republic of China; yixing@shao.ac.cn}

\author[0000-0003-1984-3852]{Zhongxiang Wang}
\affiliation{Department of Astronomy, School of Physics and Astronomy, Key Laboratory of Astroparticle Physics of Yunnan Province, Yunnan University, Kunming 650091, People's Republic of China; zhangpengfei@ynu.edu.cn; wangzx20@ynu.edu.cn}
\affiliation{Key Laboratory for Research in Galaxies and Cosmology, Shanghai Astronomical Observatory, Chinese Academy of Sciences, 80 Nandan Road, Shanghai 200030, People's Republic of China; yixing@shao.ac.cn}

\begin{abstract}
	Motivated by the newly discovery of a spin period 1.83\,ms
pulsar J1835$-$3259B in the globular cluster (GC) NGC~6652, 
	we analyze the \gr\ data obtained with the Large Area Telescope (LAT)
	onboard {\it Fermi Gamma-ray Space Telescope (Fermi)} for the GC 
and detect the pulsations of this millisecond pulsar (MSP) at 
a 5.4$\sigma$ confidence
level (the weighted H-test value is $\sim 41$).  From timing analysis of the
	data, a pulse profile that is similar to the radio one is 
established.
	We thus consider that we have detected the \gr\ emission of the MSP, 
	and discuss the implications. Based on our analysis results and
	different studies of the sources 
	in the GC, the observed \gr\ emission from the GC could mainly arise 
	from this MSP, like the previous two cases the GCs NGC~6624 and 
	NGC~6626. Assuming this is the case, the pulsar, at the GC's 
9.46\,kpc distance and having a spin-down luminosity of 
$\leq 4.3\times 10^{35}$\,erg\,s$^{-1}$,
	would have a \gr\ luminosity of 
	$\simeq (5.04\pm0.44)\times 10^{34}$\,erg\,s$^{-1}$ and 
a \gr\ efficiency of $\gtrsim 0.12$. 
\end{abstract}


\section{Introduction}
\label{Intro}
Globular clusters (GCs) are spherical ensembles of old stars that constitute
as an important part of our Galaxy. They contain $\sim10^5$ stars within 
themselves and
most of them have typical ages of greater than $10^{10}$ year 
(e.g., \citealt{Harris1996}).
Owing to the high stellar densities and thus frequent dynamical interactions 
in their cores, GCs are the natural sites to form compact low-mass X-ray 
binaries, whose further evolution resulting in formation of millisecond pulsars
(MSPs; \citealt{bv91}). Therefore GCs can be considered as 
factories of MSPs. This scenario is supported by the observational facts that
thus far 257 MSPs have been detected in 36 GCs within $\sim$20~kpc
of the Galactic center\footnote{http://www.naic.edu/$\sim$pfreire/GCpsr.html}
and these GC MSPs constitute a fraction of $\sim$50\% of known MSPs
in our Galaxy \citep{man+05}.

Since the successful launch of the \emph{Fermi Gamma-ray Space Telescope (Fermi)} and use of the Large Area Telescope (LAT) onboard \citep{Atwood2009},
the first case of \gr\ detection of GCs, the 47~Tuc, was reported
by \citep{Abdo2009}, then followed with that of Terzan~5 by \citet{Kong2010}.
Now approximately 40 GCs have been reported to have detectable \gr\ emission
\citep{1fgl2010,abd+10,Tam2011a,Zhou2015,Zhang2016,Llo+2018,Men+2019,Abdollahi2020,yua+2022}. 
It is generally considered that the \gr\ emission of GCs primarily arises
from the MSPs contained within them.
This consideration has been well supported by the discoveries of pulsed \gr\ 
emission of an individual pulsar that dominates the detected \gr\ emission
of a whole GC, specifically PSR~J1823$-$3021A in NGC~6624 \citep{fre+11}
and B1821$-$24 in NGC~6626 (or M28; \citealt{wu+13,joh+13}). Also not only
the \gr\ luminosities of the GCs but also their \gr\ spectra have been 
analyzed to show that the emission is consistent with that arising from 
a number of MSPs in each GCs \citep{abd+10,hui+11,dcn19,lcb18,zzy+20,son+21,wu+22}.

\begin{table}
\begin{center}
\caption{Likelihood analysis results}
\begin{tabular}{cccc}
\hline\hline
Models & \multicolumn{3}{c}{Parameter values} \\
\hline
LP& $\alpha$ & $\beta$ & $E_b$ (GeV) \\
	& 2.051(98) & 0.400(69) & 1.158(45)  \\
 & 2.23(12)$^\divideontimes$ & 0.29(09)$^\divideontimes$ & 1.16$^\divideontimes$ \\
\hline
PLEC & $\Gamma$ & $b$ & $E_c$ (GeV) \\
 & 1.544(95) & 1.01(14)&3.00(33)  \\
\hline
\end{tabular}
\label{tab:par}
\end{center}
	{$^\divideontimes$4FGL-DR3 values for the LP model, while no errors
	(numbers in parentheses) were given for $E_b$. }
\end{table}

Recently, \citet{gau+2022} reported the discovery of a 1.83~ms MSP, 
PSR J1835$-$3259B, in a near-circular orbit of 28.7~hr within the GC
NGC~6652. This MSP is the second one found in this GC, while the
first is PSR~J1835$-$3259A that has a spin period of 3.89~ms and is in a wide 
binary with orbital period 9.25 day \citep{dec+2015}.
NGC~6652 also shows detectable \gr\ emission, with the counterpart named 
J1835.3$-$3255 in the first \fermi\ LAT source catalog (1FGL; 
\citealt{abd+10}) and latter J1835.7$-$3258 in the fourth catalog (4FGL;
\citealt{4fgl-dr3}). Based on its \gr\ spectrum and using a source distance
of 10\,kpc \citep{Harris1996}, \citet{wu+22} estimated a number of 1--7 MSPs
in it. Motivated by these and given the detailed ephemeris for 
PSR J1835$-$3259B provided in \citet{gau+2022}, we carried out timing analysis 
of the \fermi-LAT data for NGC~6652. We have been able to detect
\gr\ pulsations of PSR J1835$-$3259B. Here we report the analysis and results.


\section{Analysis and Results}
\label{sec:lat-data}

\subsection{\fermi-LAT Data and source model}
\label{sec:model}

We selected the \emph{Fermi}-LAT Pass 8 \emph{Front+Back} events
(evclass = 128 and evtype = 3) in the
energy range of 0.1-500.0 GeV within a $20^\circ\times20^\circ$ region of 
interest (RoI) centered at NGC~6652
(R.~A.~=~$\rm18^h35^m45\farcs502$, decl.~=~$-32^{\circ}58^{'}15\farcs621$).
The time range of the data was from
2008 August 04 16:29:16.8 to 2022 June 15 22:12:11.0 (UTC).
We excluded the events with zenith angles $>90^{\circ}$
to avoid the contamination from the Earth's limb and those with quality flags 
of `bad' (by the expression
DATA\_QUAL$>$0~\&\&~LAT\_CONFIG = 1),
and the instrumental response function `P8R3 SOURCE V3' was used. 
Thus high-quality data in good time 
intervals were selected. In the analysis, the version of 
Fermitools--2.0.19 packages were used.

Based on the Data Release 3 of 4FGL \citep[4FGL-DR3;][]{4fgl-dr3}, which was 
constructed from 12-yr of LAT data, a model file was
made with the script make4FGLxml.py.\footnote{https://fermi.gsfc.nasa.gov/ssc/data/analysis/user/}
The model file included the spectral parameters of the catalog sources 
within 25$^{\circ}$ of NGC 6652, and their spectral forms provided in 
4FGL-DR3 were used.
We set free the parameters of flux normalizations and spectral shapes for the 
sources within 5$^{\circ}$. Other free parameters included
the normalizations of the sources within $5^{\circ}-10^{\circ}$, those 
outside $10^{\circ}$ but identified with $Variablility\_Index \geq 72.44$
(i.e., variable sources), and the Galactic and extragalactic diffuse 
emission components.
All other parameters were fixed at their values provided in 4FGL-DR3.
\begin{figure}
\centering
\includegraphics[angle=0,scale=0.48]{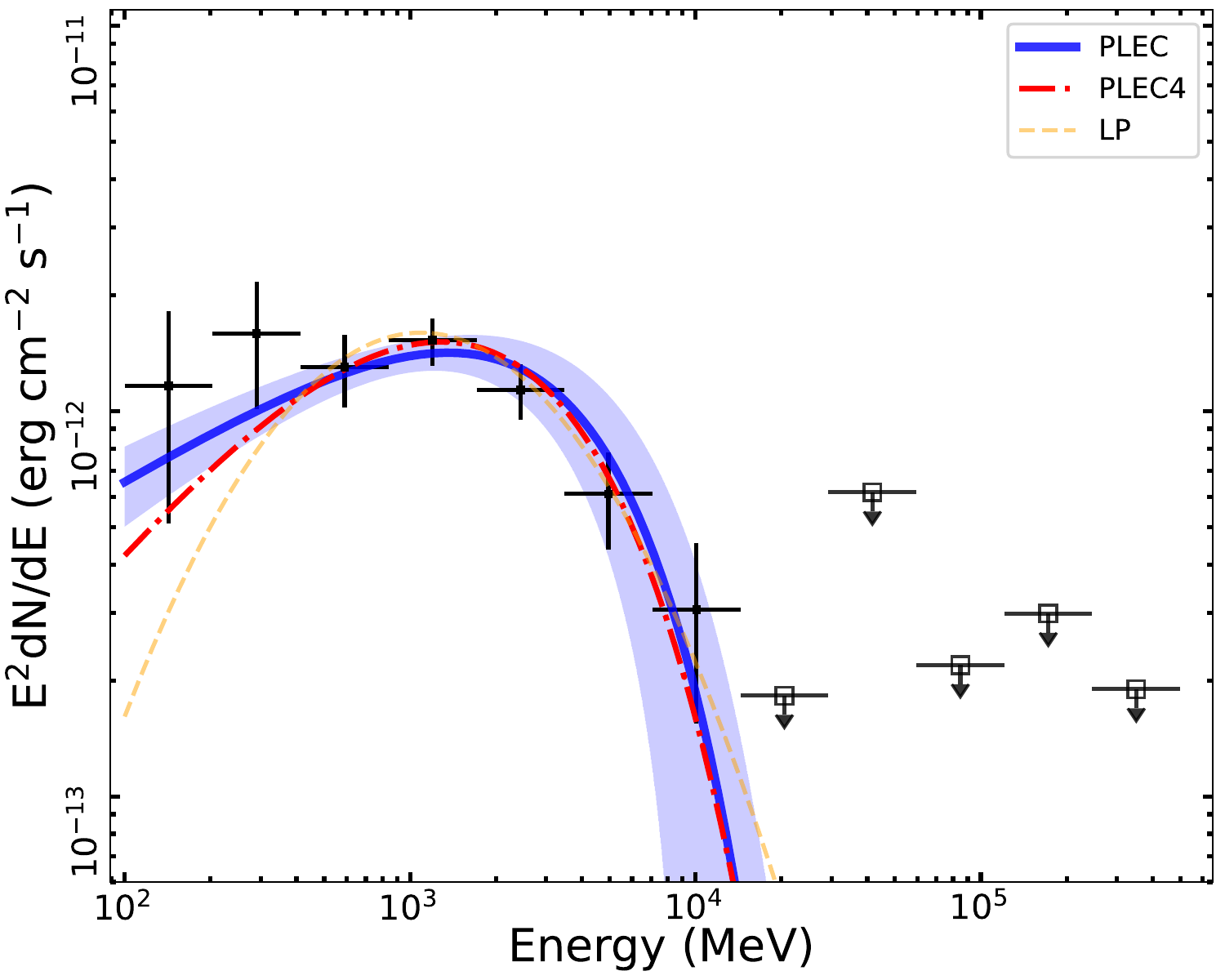}
\caption{\gr\ spectrum in 0.1--500 GeV obtained from the data for NGC~6652. 
The best-fit LP and PLEC models are shown as yellow dashed and blue solid 
lines respectively. For the latter, its error range is also shown as the 
	light blue region. A testing PLEC4 model is shown as the red 
	dash-dotted line (cf., Section~2.2)}.
\label{fig:spec}
\end{figure}

\subsection{Data analysis and spectrum extraction}

A binned maximum likelihood analysis was performed to the whole LAT 
data, where the log-parabola (LP) model provided in 4FGL-DR3 for NGC~6652,
$\frac{d N}{d E}=N_{0}(\frac{E}{E_b})^{-[\alpha+\beta \log (E / E_b)]}$, was
used.
The obtained best-fit parameter values are given in Table~\ref{tab:par}, 
which are in agreement with those given in
4FGL-DR3. The corresponding flux in 0.1--500.0 GeV was
$(4.07\pm0.74)\times10^{-9}$~photon~cm$^{-2}$~s$^{-1}$ and
test statistic (TS) value was $\simeq$212.
Because we consider that the \gr\ emission likely arises from the MSPs,
we then used the model of a power-law with an exponential cutoff (PLEC), 
$\frac{d N}{d E}=N_{0}(\frac{E}{E_{0}})^{-\Gamma} \exp [-(\frac{E}{E_{c}})^{b}]$, which is typical for describing pulsars' \gr\ emission.
Performing the likelihood analysis, the best-fit parameter values were obtained 
and are given in Table~\ref{tab:par}.
The corresponding flux in 0.1--500\,GeV was 
$(5.71\pm0.84)\times10^{-9}$~photon~cm$^{-2}$~s$^{-1}$ and
TS value was $\sim$208. As this TS value is only slightly smaller than
that from the LP model, the two models (shown in Figure~\ref{fig:spec}) 
should be considered
equally good for fitting the \gr\ emission of NGC~6652.
As in 4FGL-DR3, a new parameterization was developed for fitting
emission of pulsars, we tested the model, PLEC4 (see \citealt{4fgl-dr3} for
details), and the results were similar to that from PLEC 
(cf., Figure~\ref{fig:spec}).
\begin{figure*}
\centering
\includegraphics[angle=0,scale=0.45]{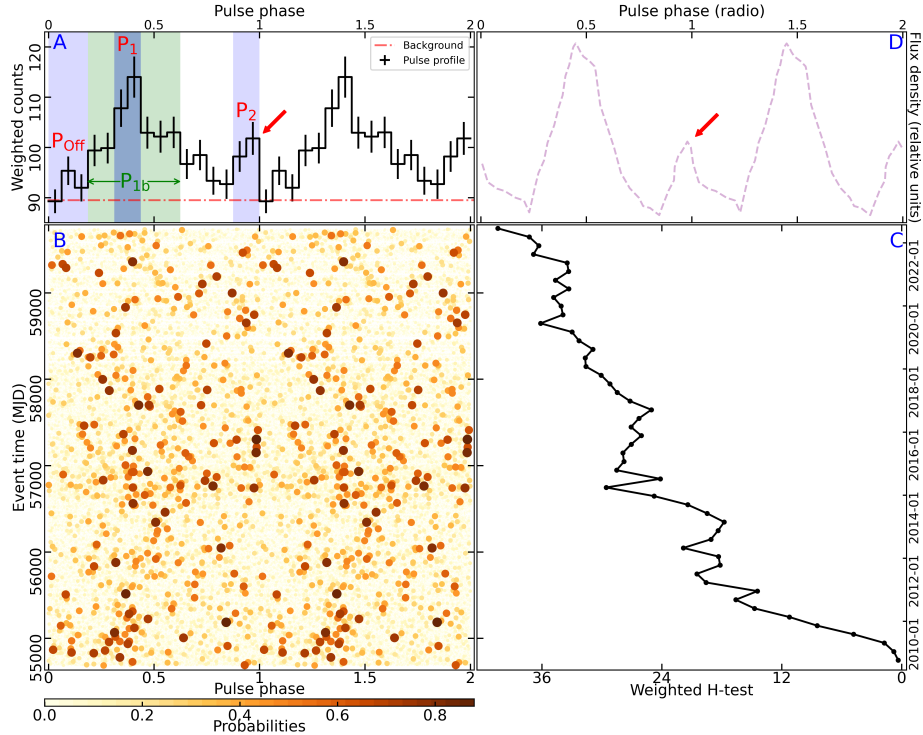}
\caption{Timing analysis results for PSR J1835$-$3259B. Panel A: the integrated 
	weighted \gr\ pulse profile (16 phase bins). Panel B: two-dimensional 
	phaseogram (16 phase bins), where the bottom color bar
	indicates the weights of the photons
	(with the largest being 88\%).
	Panel C: H-test values over the span of the data.
	Panel D: schematic radio pulse profile of 
	PSR J1835$-$3259B drawn from that reported in \citet{gau+2022} for
	comparison. Red arrows indicate the mini-peak seen in both 
	$\gamma$-rays and radio.}
\label{fig:fold}
\end{figure*}

We adopted the PLEC model in the following analysis, and the obtained
best-fit parameters were saved in the source model.
Using the best-fit parameters of the PLEC, the integrated energy flux was 
calculated to be $(4.71\pm0.39)\times10^{-12}$ ergs cm$^{-2}$ s$^{-1}$,
and considering the newly reported source distance 9.46$\pm0.14$\,kpc 
for the GC \citep{bv21}, 
the \gr\ luminosity (assuming isotropic emission)
is $L_\gamma=(5.04\pm0.44)\times10^{34}$\,erg\,s$^{-1}$.

Based on the updated source model, in which the parameters of spectral shapes 
for all the sources were fixed at their best-fit values obtained 
above and the normalizations of the sources 
within 10$^{\circ}$ of the target and the two background components were set
as free parameters, we performed spectral analysis.
The 0.1--500.0 GeV energy range was divided into 12 equal 
logarithmically-spaced energy bins. A spectrum of the whole LAT data was 
extracted by performing
the maximum likelihood analysis to the data in each energy bins.
The obtained spectrum is shown in Figure~\ref{fig:spec}, for which
we kept the flux data points with TS values $\geq$4 and showed 
95\% flux upper limits otherwise. 
This spectrum is relatively well described by the PLEC model,
whose $\Gamma$ is within and $E_c$ is slightly higher than the respective
ranges determined from the spectra of 104 \gr\ MSPs \citep{wu+22}. 

\subsection{Timing Analysis}


We selected the events within an aperture radius of 6 degrees 
(the 68\% containment angle at 0.1 GeV\footnote{https://www.slac.stanford.edu/exp/glast/groups/canda/lat\_Performance.htm}) 
in 0.1--500\,GeV band, and assigned weights to them with their probabilities of 
originating from the target (using the \fermi\ tool {\tt gtsrcprob}). 
Pulse phases were assigned to the weighted photons based on the given 
ephemeris \citep{gau+2022} by employing Tempo2 \citep{hem06}
with \fermi~plug-in \citep{rkp+11}. The two-dimensional phaseogram
and folded pulse profile in 16 phase bins are plotted in 
Figure~\ref{fig:fold} (Panel B and A respectively), where
the uncertainties of the counts in each bins were estimated using the method 
provided in \citet{2pc}.
The pulse profile shows a peak at phase $\sim$0.31--0.44,
as the results of the photons with large weights (high probabilities) 
mostly located within phase 0.3--0.5,
and also a mini-peak at phase $\sim$0.94--1.0. A high similarity to 
the shape of the radio pulse profile
(Panel D of Figure~\ref{fig:fold}, which was approximately drawn 
based on that reported
in \citealt{gau+2022}) is thus seen. Following \citet{2pc}, we estimated the
background counts (diffuse emission plus contributions from the nearby 
sources), and the value was 89.5, equal to the lowest one (89.3$\pm$2.4)
of the pulse bins.

We applied the H-test statistic to the weighted 
photons \citep{jb2010,k11}, and the cumulative H-test value curve over 
the time span of the LAT data is shown in Panel C of Figure~\ref{fig:fold}.
The H-test value of the whole data was $\simeq 41$, corresponding to a 
$p$-value of $7.5\times10^{-8}$ ($\simeq 5.4\sigma$).

\subsection{Phase-resolved analysis}

Given the established pulse profile, we first performed the 
likelihood analysis to the data of the P$_1$ and P$_2$ phase ranges shown in 
Figure~\ref{fig:fold}, for further investigation of the 
emissions of the pulse peaks.  Using the PLEC model, the fits to
the main pulse peak (P$_1$) and the mini-peak (P$_2$) were obtained.
The best-fit parameters are consistent
with those obtained from the whole data but with large uncertainty ranges,
since the TS values were $\simeq 119$ and $\simeq 33$ respectively.

We also performed the analysis to the data of the P$_{\rm 1b}$
and P$_{\rm off}$ phase ranges (Panel A of 
Figure~\ref{fig:fold}), for checking the contributions of the major pulsed 
emission and the possible off-pulse emission respectively. 
Likelihood analysis resulted in TS values of 189 and 1 respectively. The 
flux upper limit (95\%) for the latter was 
$\simeq 6.0\times 10^{-10}$\,photon\,cm$^{-2}$\,s$^{-1}$ (or 
$\sim 5.3\times 10^{33}$\,erg\,s$^{-1}$ at the GC's distance and assuming
the PLEC parameters of the MSP).

\section{Discussion}
\label{sec:dis}

By using the ephemeris given by the radio detection of PSR J1835$-$3259B, 
we have analyzed the \fermi-LAT data collected for NGC~6652 for approximately 
14 years and detected the pulsations of this newly discovered MSP
at a 5.4$\sigma$ confidence level. The high similarity of 
the \gr\ pulse profile with the radio one also strongly supports our 
detection. This pulsar has a maximum spin-down rate $\dot{P}$ of 
6.65$\times 10^{-20}$ \citep{gau+2022}, which implies a spin-down luminosity
of $\dot{E}\leq 4.3\times 10^{35}$\,erg\,s$^{-1}$. The \gr\ efficiency
would be $\eta= L_{\gamma}/\dot{E}\gtrsim 0.12$ when we assign the observed
$L_{\gamma}$ to the MSP. The $\eta$ value
is in line with those of other MSPs (e.g., \citealt{wu+22}), again supporting
the detection. 

\citet{wu+22} have estimated 1--7 MSPs, with 4 MSPs as the likely case, 
in NGC~6652.  However, the background light curve matches the pulse 
phase bin of the lowest counts and the results from the phase-resolved analysis
indicate that the major pulsed part contributes to most of the observed \gr\ 
emission from the GC, strongly suggesting little contributions from other 
sources. We note that the first MSP J1835$-$3259A had
no reported $\dot{P}$ (set to zero in \citealt{dec+2015}), which thus likely
contributes little to the observed \gr\ emission.  
\citet{pad+21} reported a candidate transitional MSP (tMSP) in the GC, which
has an X-ray luminosity of $1.8\times 10^{34}$\,erg\,s$^{-1}$. If this 
candidate is in a sub-luminous disk state or a pulsar state, 
its \gr\ luminosity would be $\sim 5\times 10^{34}$\,erg\,s$^{-1}$ 
(i.e., $\sim L_{\gamma}$)
or much higher respectively (based on the X-ray--to--$\gamma$-ray flux ratios
summarized for tMSPs in \citealt{mil+20}). The \gr\ detection of J1835$-$3259B 
actually challenges the identification of the candidate tMSP. Thus it is 
very likely that the \gr\ emission from the GC comes dominantly from 
PSR~J1835$-$3259B.

Intriguing questions remain to be investigated such as why 
like NGC~6624 and NGC~6626, NGC~6652 also contains one exceptionally bright
MSP. With \gr\ luminosities of $\sim$4--9$\times 10^{34}$erg\,s$^{-1}$, 
the three MSPs are among the brightest ones \citep{wu+22}. While the former two
have higher spin-down luminosities of $\sim 10^{36}$\,erg\,s$^{-1}$, these
three all have relatively high rotational power among all MSPs. Also the 
former two are isolated pulsars, but PSR~J1835$-$3259B is in a binary
likely with a Helium white dwarf companion \citep{gau+2022}, along 
the evolutionary tracks for such binaries \citep{ts99}. Further studies of
pulsars in the GCs could help our understanding, by pinning down how many
such bright MSPs are in the $\gamma$-ray bright GCs.

\begin{acknowledgments}
	We thank anonymous referee for very helpful suggestions and
	W. Wu for inspiring this work and helping with the MSP property
	comparison.
This work is supported in part by the National Key R\&D Program of China 
under grant No. 2018YFA0404204, the National Natural Science Foundation of 
China No.~12163006, the Basic Research Program of Yunnan Province 
No.~202201AT070137,
and the Foundations of Yunnan Province (202201BF070001-020).
Z.~W. acknowledges the support by the Original
Innovation Program of the Chinese Academy of Sciences (E085021002) and
the Basic Research Program of Yunnan Province No. 202201AS070005. 

\end{acknowledgments}

\bibliographystyle{aasjournal}
\bibliography{gcmsp}
\end{document}